\documentclass[fleqn,11pt]{wlscirep}

\usepackage{amsfonts}
\usepackage{amsmath}
\usepackage{amssymb}
\usepackage{revsymb}
\usepackage{mathrsfs}
\usepackage{verbatim}

\usepackage{graphicx}
\usepackage{dcolumn}
\usepackage{bm}

\usepackage{color}

\title{Giant collimated gamma-ray flashes}

\author[1]{Alberto Benedetti}
\author[1,*]{Matteo Tamburini}
\author[1]{Christoph H. Keitel}
\affil[1]{Max-Planck-Institut f\"ur Kernphysik, Saupfercheckweg 1, 69117 Heidelberg, Germany}
\affil[*]{matteo.tamburini@mpi-hd.mpg.de}

\begin{abstract}
\textbf{Bright sources of high energy electromagnetic radiation are widely employed in fundamental research as well as in industry and medicine~\cite{lightsources}. 
This steadily growing interest motivated the construction of several facilities aiming at the realisation of sources of intense X- and gamma-ray pulses~\cite{DESY, ELI-NP}. 
To date, free electron lasers and synchrotrons provide intense sources of photons with energies up to~\cite{Ullrich} 10-100~keV. 
Facilities under construction based on incoherent Compton back scattering of an optical laser pulse off an electron beam are expected to yield photon beams with energy up to 19.5~MeV and peak brilliance in the range 10$^{\boldsymbol{20}}$-10$^{\boldsymbol{23}}$ photons s$^{\boldsymbol{-1}}$ mrad$^{\boldsymbol{-2}}$ mm$^{\boldsymbol{-2}}$ per 0.1\% bandwidth \cite{ELI-NP}. 
Here, we demonstrate a novel mechanism based on the strongly amplified synchrotron emission which occurs when a sufficiently dense 
electron beam interacts with a millimetre thickness solid target. 
For electron beam densities exceeding approximately $\boldsymbol{3\times10}^{\boldsymbol{19}}\text{ cm$^{\boldsymbol{-3}}$}$ filamentation instability occurs with the self-generation of 10$^{\boldsymbol{7}}$-10$^{\boldsymbol{8}}$~gauss magnetic fields where the electrons of the beam are trapped. 
This results into a giant amplification of synchrotron emission with the production of collimated gamma-ray pulses with peak brilliance above $\boldsymbol{10}^{\boldsymbol{25}}$ photons s$^{\boldsymbol{-1}}$ mrad$^{\boldsymbol{-2}}$ mm$^{\boldsymbol{-2}}$ per 0.1\% bandwidth and photon energies ranging from 200~keV up to several hundreds~MeV. 
These findings pave the way to compact, high-repetition-rate (kHz) sources of short (30~fs), collimated (mrad) and high flux ($\boldsymbol{>10^{\boldsymbol{12}}}$ photons/s) gamma-ray pulses.}
\end{abstract}

\begin{document}

\flushbottom
\maketitle
\thispagestyle{empty}



Thanks to their short duration, high brightness and wide energy tunability, conventional synchrotron sources are extensively used for interdisciplinary research in physics, material science, chemistry and biology \cite{lightsources, 0953-4075-38-9-022}.
However, these facilities are still large and limited in the attainable photon energy and intensity.
Recently, due to considerable improvements in strong-field laser technology, alternative methods have also been considered to produce brilliant sources of photons in the $\gamma$-ray region \cite{RevModPhys.84.1177}. 
In particular, Thomson- and Compton-based sources have been tested in proof-of-principle experiments by crossing a super-intense laser pulse with a picosecond duration relativistic electron beam from a conventional linear electron accelerator \cite{PhysRevLett.113.224801,PhuocK.2012}.


Here we show that a 2~GeV electron beam with milliradiant divergence and density above about $3\times10^{19}$~cm$^{-3}$ generates a magnified emission of synchrotron radiation while travelling across a conductor as thick as 0.5~mm.
Counter-intuitively, most of the beam energy is rapidly and efficiently converted into a $\gamma$-ray flash whose brilliance is dominated by synchrotron emission rather than by bremsstrahlung emission. 
The physical picture of this phenomenon goes as follows. 
When the electron beam enters the target, the free electrons of the target move towards the opposite direction with respect to the beam velocity in order to neutralise its current within the relaxation time-scale $\tau_e=1/4\pi\sigma$, where $\sigma$ is the conductivity of the medium\cite{2001PhPl....8.1441F}. 
For typical conductors\cite{047141526X} $\sigma \approx 10^{17}$ s$^{-1}$. Thus, after about $10^{-18}$~s, the configuration of two overlapping counter-propagating currents is established and, because of the collective plasma dynamics, the electron beam becomes unstable to small electromagnetic fluctuations\cite{PhysRevLett.2.83}. 
In view of the parameters considered in our work, filamentation is the fastest growing instability 
and takes place over a time-scale of the order of $\tau_F=1/\delta_F\sim10^{-13}$~s, where $\delta_F\approx\sqrt{4\pi e^2n_b/\gamma_bm_e}$ is the instability growth rate, $e$ and $m_e$ are the electron charge and mass, while $n_b$ and $\gamma_b$ are the electron beam density and average Lorentz factor, respectively\cite{MillerChargedBeams}. 
As a result, the electron beam splits into small filaments parallel to the beam velocity with radius of the order of the target plasma skin depth $c/\omega_e$, where $c$ is the speed of light in vacuum, $\omega_e=\sqrt{4\pi e^2n_e/m_e}$ the plasma frequency of the target, and $n_e$ the number density of the free target electrons\cite{PhysRev.137.A1083}. 
During the initial phase of the instability, the free electrons of the target cannot neutralise the beam current within each filament, and the high-energy electrons of the beam travel across, and are confined by, the large self-generated electromagnetic fields\cite{PhysRevLett.96.105008}.  
In the second phase of the instability, filaments attract each other and tend to merge\cite{PhysRevLett.31.1390,AW2}, therefore further increasing the strength of the self-generated electromagnetic fields. The simultaneous occurrence of ultra-relativistic electrons and overlapping super-strong electromagnetic fields results into a giant emission of synchrotron radiation. 

\begin{figure}[ht] 
\centering
\includegraphics[width=0.95\linewidth]{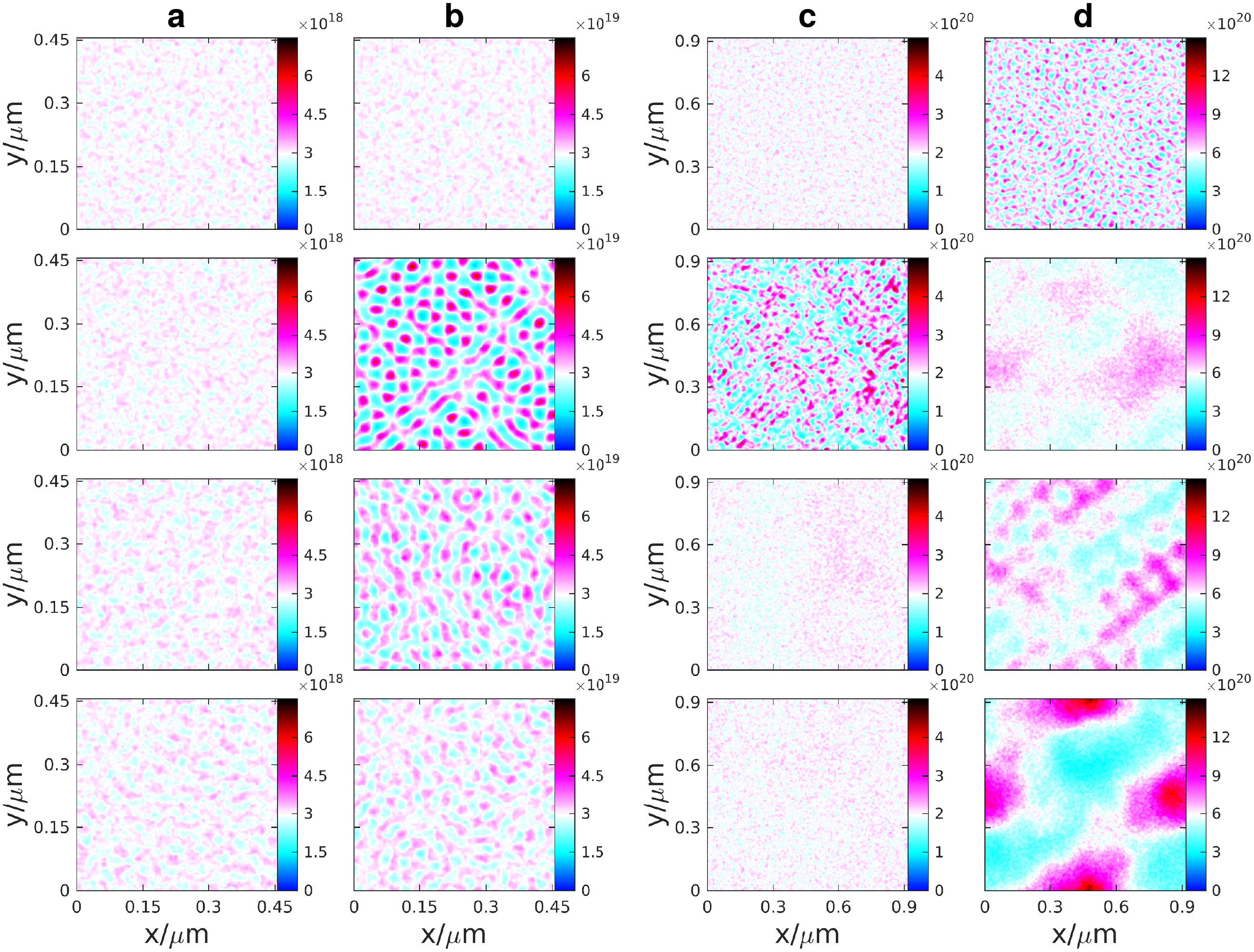}
\caption{{\bf The electron beam density}. Evolution of the electron beam number density distribution obtained from 3D PIC simulations with: 
\textbf{a}, $n_b=3\times10^{18}$~cm$^{-3}$, \textbf{b}, $3\times10^{19}$~cm$^{-3}$, \textbf{c}, $2\times10^{20}$~cm$^{-3}$, \textbf{d}, $6\times10^{20}$~cm$^{-3}$. Plots belonging to the same row report the density distribution when the beam has travelled across the same distance $z$ inside the target. From the top to the bottom: $z=[0.2,\,0.8,\,2,\,3]\times 10^{-2}$~cm. 
Contour plots are obtained from the 3D spatial beam distribution by projecting the position of the electrons with $z$-coordinate in the interval $[0,\Delta l_z]$, $\Delta l_z$ being the spatial step, onto the $xy$-plane which is perpendicular to the initial beam velocity.  
The white band in the colorbar corresponds to the initial electron beam density $n_b$. 
Note that the scale in \textbf{a,b} is different from \textbf{c,d}.
}
\label{fig:eb_dens}
\end{figure}


In order to investigate the interaction of a high-density ultra-relativistic electron beam with a solid conductor quantitatively, we performed fully three-dimensional (3D) particle-in-cell (PIC) simulations with synchrotron and bremsstrahlung emission included (see Methods). 
In all our simulations, the electron beam has initial Lorentz factor $\gamma_b=4\times10^3$, 6\% energy spread, and 0.8 mrad angular divergence. Note that similar parameters have already been attained in electron laser wake-field acceleration experiments\cite{PhysRevLett.117.124801,Wang2013,PhysRevLett.113.245002}. In order to demonstrate the critical importance of the electron beam density, in our simulations the initial beam density $n_b$ was varied from $3\times10^{18}$~cm$^{-3}$, which was already achieved with laser-generation techniques\cite{Malka2008}, up to $6\times10^{20}$~cm$^{-3}$. 
Within the considered density range, an electron beam with radius of about 5~$\mu$m carries a  current $I_b\lesssim4\times10^{-2}\,I_A \ll I_A$, where $I_A\approx17\,\gamma_b$ kA is the so called Alfv\'en limit\cite{MillerChargedBeams}. 
This is important because for electron beams propagating into the vacuum with $I_b\gtrsim I_A$ the magnetic field generated by the current itself would be sufficient to reverse the direction of the electron trajectories at the outer edge of the beam\cite{MillerChargedBeams}. 
For what concerns the target, here we employ metallic Strontium (atomic number $Z=38$) because of its high electrical conductivity and large radiation length. While the high electrical conductivity ensures the rapid occurrence of the conditions for the onset and growth of the filamentation instability, the large radiation length prevents a premature deterioration of the brilliance that is eventually caused by the lateral broadening of the beam due to multiple elastic scattering events, which were also included in the PIC code (see Methods for details). 


In order to visualise how the beam filamentation evolves, in Fig.~\ref{fig:eb_dens} we plot the electron beam density distribution in the $xy$-plane, which is perpendicular to the initial beam velocity $\boldsymbol{v}_b=(0,0,v_b)$, when the beam has travelled across a distance $z$ inside the target. In Fig.~\ref{fig:eb_dens}, four values of the initial electron beam density are considered: $n_b=[0.03,\,0.3,\,2,\,6]\times 10^{20}$~cm$^{-3}$. Figure~\ref{fig:eb_dens} shows that, in agreement with the theoretical instability growth rate\cite{MillerChargedBeams} $\delta_F\sim10^{13}(n_b/10^{20}\text{cm}^{-3})^{1/2}(\gamma_b/4\times10^3)^{-1/2}$~s$^{-1}$, 
for the same initial Lorentz factor $\gamma_b$ denser beams undergo filamentation earlier. However, although filaments form sooner while moving from lower to higher densities (see Fig.~\ref{fig:eb_dens}), from the quantitative point of view the above theoretical prediction for $\delta_F$ overestimates the actual growth rate of the instability to the extent that filamentation does not even occur for $n_b=3\times10^{18}$~cm$^{-3}$. 
The source of the discrepancy is well understood and is related to the electron beam transverse temperature, which slows down or even stops the instability\cite{PhysRevE.72.016403}, because it acts against the self-generated field confinement. Note that, in addition to the initial electron beam temperature, multiple scatterings of the electrons of the beam with the atoms of the target also result in an increase of the electron beam transverse temperature while the beams propagates through the target (see Methods). This explains another noticeable feature that emerges from Fig.~\ref{fig:eb_dens} and concerns the non-linear stage of the instability, i.e. when the filaments attract one another and tend to merge. In fact, while for $n_b=[0.3,\,2]\times 10^{20}$~cm$^{-3}$ filaments are disrupted by the multiple elastic scattering events before filament merging occurs (see Figs.~\ref{fig:eb_dens}b-c), for $n_b=6\times 10^{20}$~cm$^{-3}$ filaments merge and give origin to more complex density patterns (see Fig.~\ref{fig:eb_dens}d). 


\begin{figure} 
\centering
\includegraphics[width=\linewidth]{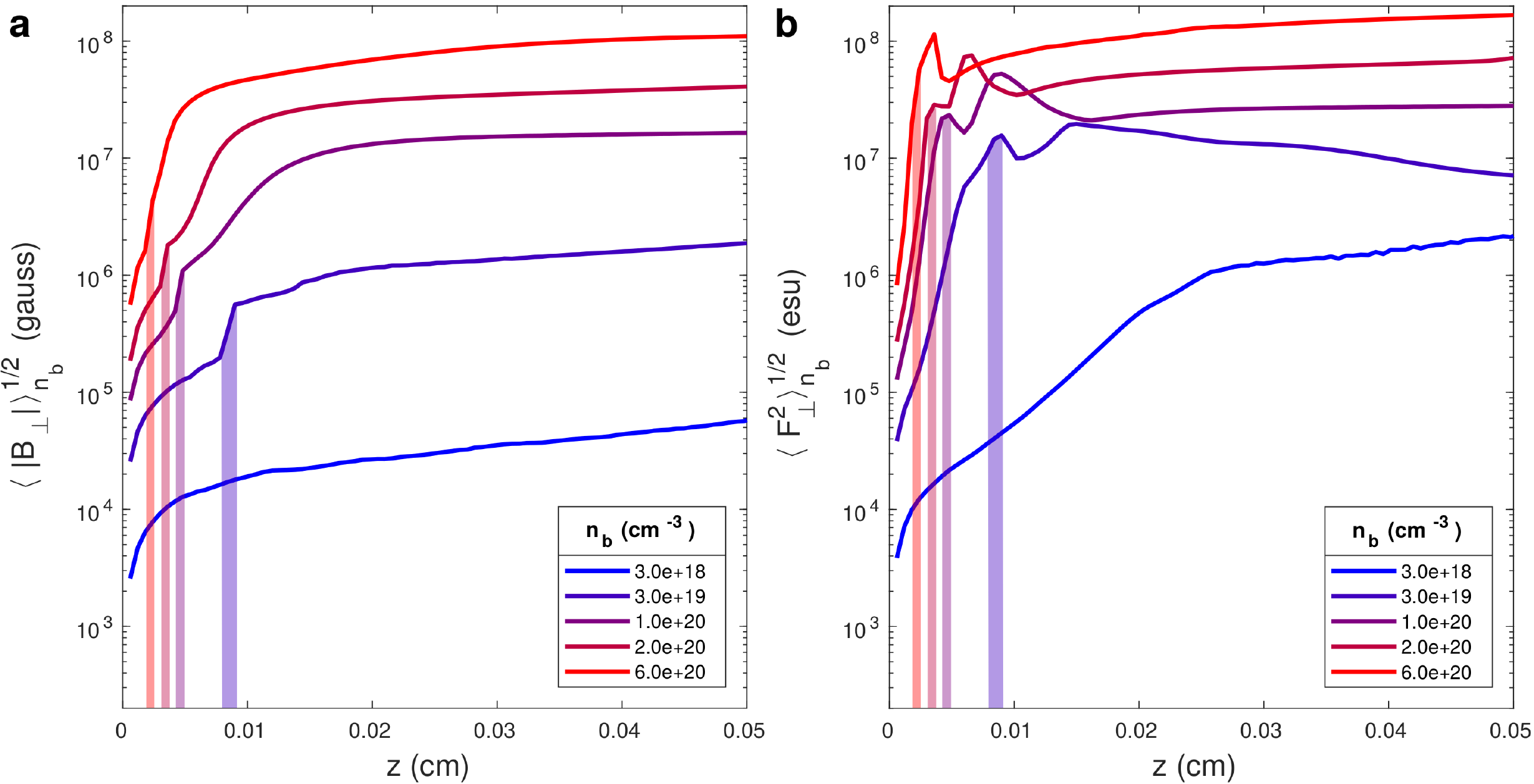}
\caption{{\bf Electromagnetic fields}. \textbf{a}, average transverse magnetic field experienced by the electrons of the beam $\langle B_\perp^2\rangle_{n_b}^{1/2}$ as a function of the target thickness for different initial beam densities. 
\textbf{b}, same as in \textbf{a} but for the average transverse electromagnetic field experienced by the electrons of the beam $\langle F_\perp^2\rangle_{n_b}^{1/2}$.  
The vertical stripes in \textbf{a} and \textbf{b} highlight the field strength amplification associated with the formation of the first filaments in the electron beam density (only for $n_b\geq3\times10^{19}$ cm$^{-3}$).}
\label{fig:BperpFperp}
\end{figure}

The filament formation process is always associated with the self-generation of strong transverse electromagnetic fields. 
In Fig.~\ref{fig:BperpFperp}a, we show the average transverse magnetic field experienced by the electrons of the beam $\langle B_\perp^2\rangle_{n_b}^{1/2}$ as a function of the target thickness. Here, $\langle B_\perp^2\rangle_{n_b}^{1/2}$ is defined as:

\begin{equation}
\langle B_\perp^2\rangle_{n_b}=\left(\int d^3\boldsymbol{r} \;n_b(\boldsymbol{r})\right)^{-1}
\displaystyle \int d^3\boldsymbol{r} \;n_b(\boldsymbol{r}) \,\Big[B_x(\boldsymbol{r})^2+B_y(\boldsymbol{r})^2\Big]\;,\label{Bperp_ave_nb}
\end{equation}
where $n_b(\boldsymbol{r})$ is the electron beam number density distribution and the integral is performed over the whole 3D computational volume. Note the steep exponential rise of $\langle B_\perp^2\rangle_{n_b}^{1/2}$ for electron beam densities $n_b\geq3\times10^{19}$~cm$^{-3}$, highlighted by the vertical bands in Fig.~\ref{fig:BperpFperp}, which is the typical signature for the onset of the filamentation instability\cite{AW2}. 
By contrast, when $n_b=3\times10^{18}$~cm$^{-3}$ this characteristic feature is absent because, as it has been shown in Fig.~\ref{fig:eb_dens}a, the occurrence of filamentation instability is hindered by the initial transverse electron beam temperature and by multiple scattering of the electrons of the beam with the atoms of the target. 
In addition to the magnetic field, a strong transverse electric field is induced by the rapid growth of the magnetic field and by the electron bunching in the filaments, which significantly affects the photon emission processes. In fact, the emission of synchrotron radiation from ultra-relativistic electrons is governed by their transverse acceleration\cite{BaierBook} $a_\perp$, which is proportional to the transverse electromagnetic field $F_\perp=\sqrt{(E_x-B_y)^2+(E_y+B_x)^2}$ (see Methods for details). In Fig.~\ref{fig:BperpFperp}b, we plot the average transverse electromagnetic field $\langle F_\perp^2\rangle_{n_b}^{1/2}$ experienced by the electrons of the beam, which is defined as:

\begin{equation}
\langle F_\perp^2\rangle_{n_b}=\left(\int d^3\boldsymbol{r} \;n_b(\boldsymbol{r})\right)^{-1}
\displaystyle \int d^3\boldsymbol{r} \;n_b(\boldsymbol{r}) \,\Big[\big(E_x(\boldsymbol{r})-B_y(\boldsymbol{r})\big)^2+\big(E_y(\boldsymbol{r})+B_x(\boldsymbol{r})\big)^2\Big]\;,\label{Fperp_ave_nb}
\end{equation}
as a function of the target thickness. During the whole initial phase of the electron beam-target interaction and the filament formation, $F_\perp$ grows rapidly and remains above $10^7$~esu over a distance $\Delta z\approx 0.04$~cm. 
Thus, the ultra-relativistic electrons of the beam interact with super-strong electromagnetic fields for approximately $\Delta t=\Delta z/c\approx1.3$~ps. 
For comparison, in Thomson- and Compton-based $\gamma$-ray sources the beam-laser collision period lasts only a few tens of femtoseconds, and electrons have much less time to convert their kinetic energy into photons. 
Note that, although the electron beam-target interaction lasts for $\Delta t\approx1.3$~ps, the duration of the generated $\gamma$-ray beam $\Delta t_\gamma$ is of the same order as the duration of the electron beam $\Delta t_\gamma \sim l_b/c\approx 33$~fs, where $l_b\approx10\,\mu$m is the typical duration of laser-plasma generated electron beams. In fact, here electrons remain ultra-relativistic throughout the beam-target interaction, such that electrons and $\gamma$-photons travel across the target with almost the same velocity.  


\begin{figure} 
\centering
\includegraphics[width=\linewidth]{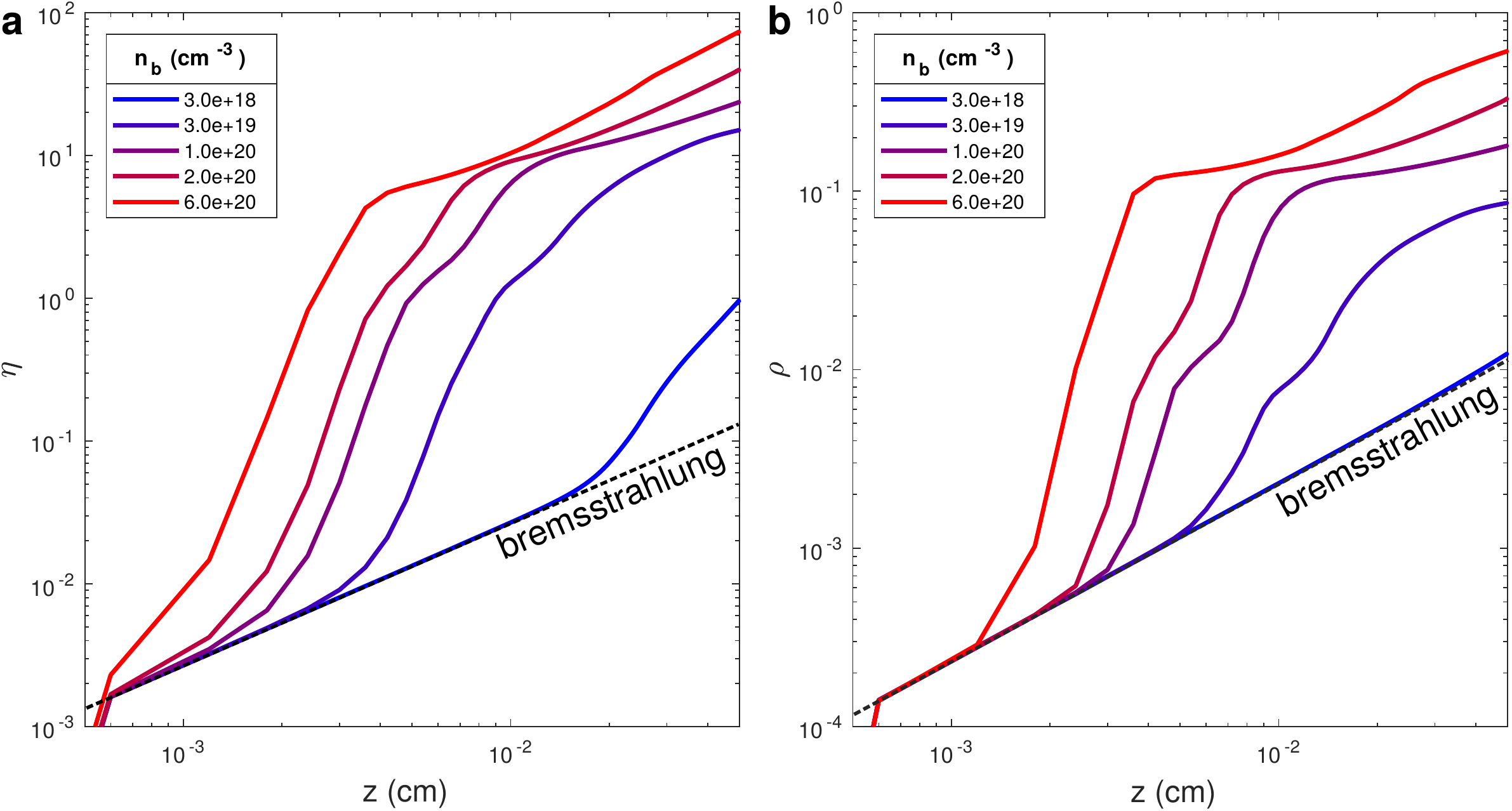}
\caption{{\bf Number of emitted photons and conversion efficiency per electron.} \textbf{a}, average number of photons emitted per electron $\eta$ as a function of the target thickness for different initial beam densities. 
\textbf{b}, the same as in \textbf{a} for the electron-to-photon energy conversion efficiency $\rho$, defined as the ratio between the photon and the initial beam total energies. In both panels the black dashed line shows the expected value for a pure bremsstrahlung emission.}
\label{fig:NgEg_norm}
\end{figure}

In Fig.~\ref{fig:NgEg_norm}a, we show the average number of photons generated per electron $\eta$ as a function of $z$. 
For $n_b=3\times10^{18}$~cm$^{-3}$, $\eta$ initially scales as $\eta\propto z$ as expected for a pure bremsstrahlung emission when $z \ll X_\text{0Sr}/d_\text{Sr}\approx4.24$~cm, where $d_\text{Sr}\approx2.54$~g/cm$^{3}$ and $X_{0Sr}\approx10.76$~g/cm$^2$ are the target mass density and radiation length of Strontium, respectively. A departure from the linear emission expected for bremsstrahlung occurs for $z \gtrsim 0.02$~cm, due to a relatively modest synchrotron emission. 
In fact, by looking at the evolution of the electron-to-photon energy conversion efficiency $\rho$, defined as the ratio between the photon and the initial electron beam total energies, we find that for $n_b=3\times10^{18}$~cm$^{-3}$ there is no appreciable deviation from the bremsstrahlung prediction\cite{RevModPhys.46.815,SegreNucleiParticles} $\rho(z)=1-\exp(-zd/X_0)\sim zd/X_0$ (see the blue curve in Fig.~\ref{fig:NgEg_norm}b). 
The situation changes drastically for $n_b\geq3\times10^{19}$~cm$^{-3}$. 
During the initial phase of the instability, the rise in both the number of emitted photons per electron $\eta$ and the electron-to-photon energy conversion efficiency $\rho$ departs swiftly from linearity because of the fast amplification of the self-generated electromagnetic fields as shown in Fig.~\ref{fig:BperpFperp}. 
At later times, when the field growth is slower, $\eta$ and $\rho$ continue to increase steadily but at a moderate pace. 
Remarkably, after crossing just a small fraction of the target radiation length, each electron produces several tens of synchrotron $\gamma$-rays losing from 10\% up to 60\% of its energy in the process (see Figs.~\ref{fig:NgEg_norm}a-b). 
This can be easily explained recalling that for synchrotron radiation the total energy emitted into high energy photons (average photon energy) scales quadratically (linearly) with the transverse field $F_\perp$ (see Methods), and that $F_\perp$ is exponentially amplified as the onset of filamentation occurs (see Fig.~\ref{Fperp_ave_nb}b).

For light sources, the number of photons delivered per shot and the repetition rate are among the most critical specifications. 
In our setup, the number of photons produced by a single electron beam with radius $r_b$ and length $l_b$ after travelling a distance $z$ in the target is estimated as:

\begin{equation}\label{N_gamma}
N_{\gamma/\text{beam}}(z) \approx
n_b\,\pi\, r_b^2\,l_b\,\eta(z) =
8\times10^{10}\left(\frac{n_b}{10^{20}\;\text{cm}^{-3}}\right)
\left(\frac{r_b}{5\,\mu\text{m}}\right)^2
\left(\frac{l_b}{10\,\mu\text{m}}\right)\,\eta(z)\;,
\end{equation}
where conventional parameters are used for normalization. Thus, assuming the typical size $r_b\approx5$~$\mu$m and length (duration) $l_b\approx10$ $\mu$m ($l_b/c\approx33$~fs) of a laser-generated electron beam, and by replacing $\eta(z)$ in equation~(\ref{N_gamma}) with the numerical values attained by our 3D PIC simulations for $z=0.05$~cm (see Fig.~\ref{fig:NgEg_norm}a), we estimate that our scheme yields more than $10^{12}$ $\gamma$-photons per shot for electron beam densities approaching $10^{20}$~cm$^{-3}$.  Since the beam filamentation occurs only if the beam length is several times the target plasma skin depth, the duration of the gamma-ray pulse that is expected to be generated with our setup is limited from below. In particular, by assuming the number density of the free target electrons considered in this work, we obtain $l_b^\text{min} \gg c/\omega_e \approx 0.013~\mu$m, i.e. the pulse duration must be $l_b^\text{min} / c \gg 1/ \omega_e \approx 0.04$~fs.


Regarding the realisation of the experimental setup we put forward in this manuscript, we stress that in laser-plasma generated electron beams the size $r_b$ and length $l_b$ of the electron beam are always much smaller than the transverse size ($\sim$~cm) and width ($\sim$~mm) of the target. Thus, only a volume of the order of $\pi r_b^2l_b\sim800$~$\mu$m$^3$ within the target is affected by the passage of the electron beam, and at most for a few tens of femtoseconds. As a consequence, ionisation and heating due to the electron beam travelling across the target cause limited damage to the target itself, such that high-repetition-rate lasers can be employed to generate the electron beam, therefore noticeably increasing the attainable average photon flux.  


\begin{figure} 
\centering
\includegraphics[width=\linewidth]{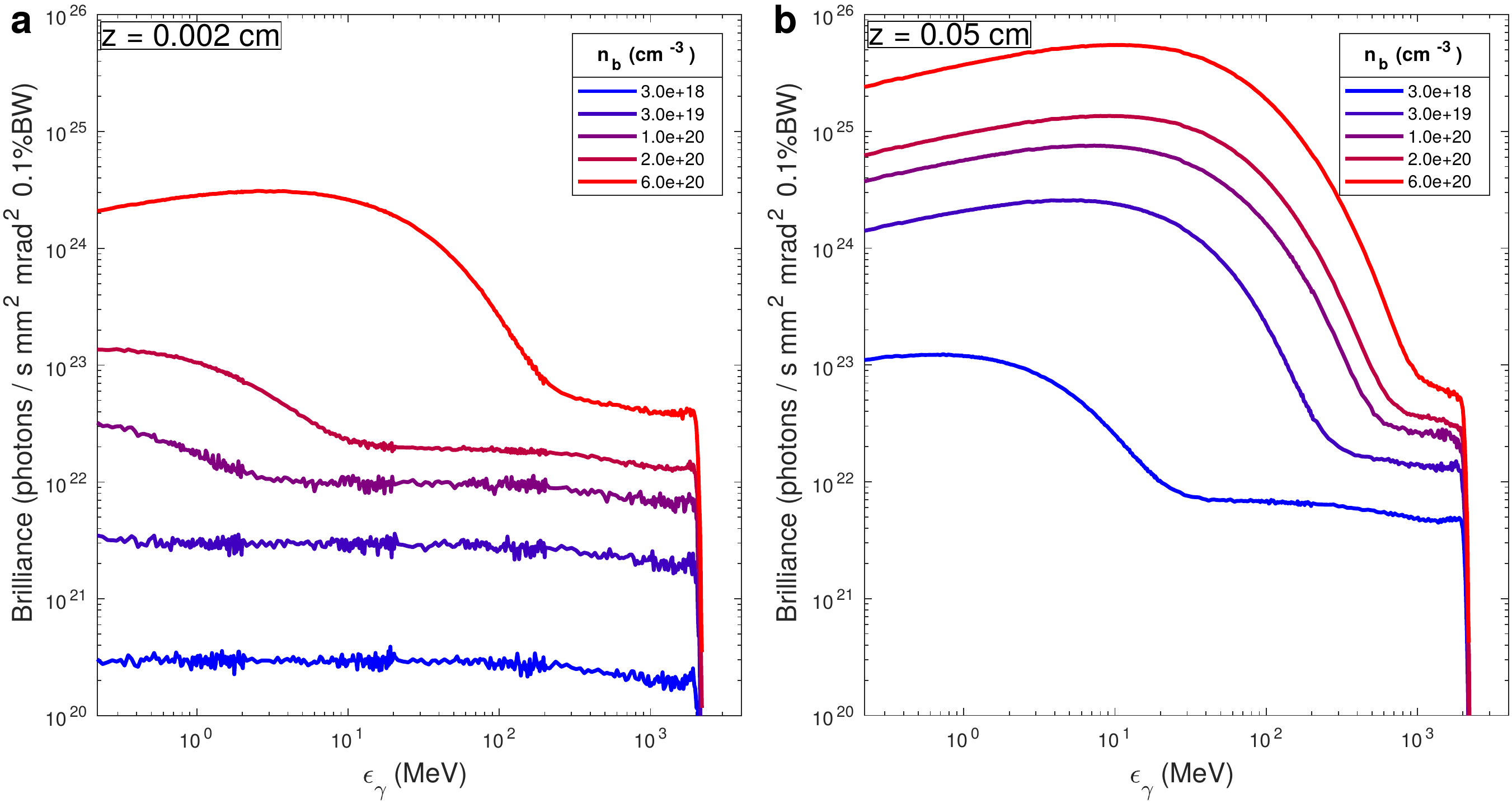}
\caption{{\bf The photon beam brilliance}. The $\gamma$-ray brilliance as a function of the photon energy $\epsilon_\gamma$ and for different initial electron beam densities. \textbf{a}, after the electron beam has traversed $z=0.002$~cm, and \textbf{b}, after the electron beam has traversed $z=0.05$~cm within the target.}
\label{fig:brilliance}
\end{figure}

Among the specifications of a light source, its brilliance is of key importance as it accounts for the flux, collimation and energy bandwidth of the generated photon beam. The brilliance is a function of the photon energy $\epsilon_\gamma$, and it is defined as the number of photons emitted per unit time (s), per unit area (mm$^2$), per unit solid angle (mrad$^2$) within an energy interval of $10^{-3}\epsilon_\gamma$ centered around $\epsilon_\gamma$. 
In Fig.~\ref{fig:brilliance}a and Fig.~\ref{fig:brilliance}b, we report the brilliance of the emitted radiation after the beam has traversed 0.002~cm  and 0.05~cm of the target width, respectively.  
For 0.002~cm target thickness and for beam densities $\leq3\times10^{19}$~cm$^{-3}$, the brilliance is dominated by bremsstrahlung emission. In fact, the brilliance is proportional to the photon energy spectrum $\epsilon_\gamma\;dN_\gamma/d\epsilon_\gamma$, where $dN_\gamma/d\epsilon_\gamma$ is the number of emitted photons per unit of photon energy. In addition, assuming only bremsstrahlung emission, $dN_\gamma/d\epsilon_\gamma$ is proportional to the bremsstrahlung differential cross section $d\sigma_\text{bs}/d\epsilon_\gamma=f(\epsilon_\gamma)/\epsilon_\gamma$, where $f(\epsilon_\gamma)$ is a function which is relatively constant over a broad range of $\epsilon_\gamma$ (see Methods). Thus, the curve of the brilliance as obtained only from bremsstrahlung emission is relatively flat, as shown in Fig.~\ref{fig:brilliance}a for beam densities $n_b \leq 3\times 10^{19}$~cm$^{-3}$. By contrast, although the traverse target thickness is only 0.002~cm, for beam densities $\gtrsim 10^{20}$~cm$^{-3}$ a prominent synchrotron emission is already visible. 
Fig.~\ref{fig:brilliance}b shows the brilliance when the target is as wide as 0.05~cm. 
For this macroscopic width, even for the lowest considered beam density $n_b=3\times10^{18}$~cm$^{-3}$ where filamentation instability is hindered by the electron beam transverse temperature, $F_\perp$ becomes large enough to generate synchrotron radiation with enhanced brilliance in the lower energy part of the photon spectrum (see the blue curve in Fig.~\ref{fig:brilliance}b). In fact, the number of emitted photons per electron after 0.05~cm of propagation increases from the bremsstrahlung prediction of $\approx0.1$ to $\approx 1$ due to synchrotron emission (see the blue curve in Fig.~\ref{fig:NgEg_norm}a). However, the energy of the emitted photons remains much smaller than the electron energy, such that the average energy loss per electron after 0.05~cm of propagation is about 1\% of the initial electron energy, and basically coincides with the bremsstrahlung prediction (see the blue curve in Fig.~\ref{fig:NgEg_norm}b). Finally, for the highest density considered $n_b=6\times10^{20}$~cm$^{-3}$, the peak brilliance increases beyond $10^{25}$ s$^{-1}$ mrad$^{-2}$ mm$^{-2}$ per 0.1\% bandwidth for $\gamma$-rays with energy up to few hundreds MeV (see the red curve in Fig.~\ref{fig:brilliance}b), and each electron converts about 60\% of its initial kinetic energy into high-energy photons (see the red curve in Fig.~\ref{fig:NgEg_norm}b).

In summary, we have demonstrated that giant $\gamma$-ray flashes can be produced by means of filamentation-unstable electron beams travelling across a millimetre thickness solid conductor. 
To enter this regime, the electron beam needs to have a few GeV energy, milliradiant divergence and number density higher than about $3\times10^{19}$~cm$^{-3}$. 
Once the above conditions are satisfied, ultra-relativistic electrons interact extensively ($\sim$ps) with strong (10$^7$-10$^8$~esu) self-generated electromagnetic fields and, within a small fraction of the target radiation length, convert from 10\% up to 60\% of their initial energy into high-energy photons by the emission of synchrotron radiation. 
For energies up to a few hundreds MeV, the $\gamma$-ray brilliance is dominated by synchrotron emission, and reaches peak values of the order of $10^{24}$-$10^{25}$ photons s$^{-1}$ mrad$^{-2}$ mm$^{-2}$ per 0.1\% bandwidth. One single electron beam, with radius $r_b=5~\mu$m and length $l_b=10~\mu$m, gives origin to a short $\sim33$~fs $\gamma$-ray pulse with $\gtrsim10^{12}$ photons. Since kHz repetition rate lasers can be employed to generate the electron beams, the attainable average photon flux is of the order of $10^{15}$ $\gamma$-photons per second. Our method paves the way for a new generation of compact synchrotron light sources which allow for high sensitivity investigations in QED such as light-by-light scattering, birefringence and dichroism of the polarised vacuum, and the catalytic generation of electron-positron showers and cascades by high-energy photons as well as the production of large amounts of excited nuclei and isotopes, which is critical for accurate studies of the nuclear structure and of nucleosynthesis processes.

\section*{Methods}\label{methods}

\subsection*{Numerical simulations}

In order to investigate the interaction of an ultra-relativistic electron beam with a solid conductor, we embedded routines accounting for multiple scattering, bremsstrahlung and synchrotron emission into the PIC code EPOCH \cite{Arber:2015hc}.

\subsubsection*{Setup}

We initialise the electron beam with Lorentz factor $\gamma_b=4\times10^3$ and average momentum $\langle \boldsymbol{p}\rangle=(\langle p_x\rangle,\langle p_y\rangle,\langle p_z\rangle)=(0,0,m_e\gamma_bv_b)$, where $v_b=c(\gamma_b^2-1)^{1/2}/\gamma_b$ is the beam longitudinal velocity. 
Both longitudinal and transverse momenta distributions are Gaussian centred on the corresponding mean values with standard deviations $\langle p_{z}^2\rangle^{1/2}=0.03\,\gamma_bm_ec$ and $\langle p_{x}^2\rangle^{1/2}=\langle p_{y}^2\rangle^{1/2}=3\,m_ec$, respectively. Initially, the electrons of the beam are uniformly distributed in the physical space with density $n_b$. 

The computational box has periodic boundary conditions in all directions, and its volume is $0.46\times0.46\times0.46~\mu\text{m}^3$ with $80\times80\times80$ grid points for $n_b=[0.03,0.3,1]\times10^{20}$~cm$^{-3}$, while the volume is $0.92\times0.92\times0.92~\mu\text{m}^3$ with $160\times160\times160$ grid points for $n_b=[2,6]\times10^{20}$~cm$^{-3}$. For the two higher electron beam densities, we employ a larger 3D computational box in order to prevent spurious effects induced by the boundary conditions when  the filaments merge. Note that in all the simulations, the spatial resolution is $\Delta l_{[x,y,z]}\approx5.8\times10^{-3}$~$\mu$m, i.e. several times smaller than the target plasma skin depth, such that the filament formation and dynamics is accurately modelled.

The selected material for the target is Strontium (Sr): atomic number $Z=38$, mass number $A=87.62$, mass density $d=2.54$~g/cm$^3$, radiation length\cite{Olive:2016xmw} $X_0=10.76$~g/cm$^2$. 
The target is parametrised by its degree of ionisation $\xi$, implying that only a fraction $\xi$ of all target electrons is free while the other fraction $1-\xi$ is bound to the nuclei. Bound electrons and nuclei are described as a single particle species with effective charge $Z_\text{eff}=\xi Z$ to which we refer as ions. 
The ion density is $n_i=N_Ad/A$, where $N_A$ is the Avogadro's constant, whilst the ion bulk momentum is $\boldsymbol{p}_i=0$. 
In order to take into account that the beam length ($l_b\sim10~\mu$m) is much smaller than the target width ($\Delta z\sim\text{mm}$), such that the electrons of the beam interact with different atoms of the target during the propagation, we assign an ideally infinite mass to the ions. 
The results presented in this work are obtained assuming $\xi^*=0.24$. 
However, this specific choice is not prejudicial to the generality of our conclusions. In fact, we also performed PIC simulations with $\xi_\text{low}=0.08$ and $\xi_\text{high}=0.75$, which lead to similar outcomes for what concerns the onset of filamentation, the amplification of the self-generated electromagnetic fields and, most importantly, the brilliance of the emitted radiation. 
One can estimate $\xi$ and the maximal electron beam energy lost in the ionisation processes by comparing the total beam energy with the energy required to ionize the volume of the target traversed by the electron beam. 
The total energy of the electron beam is $U_e=\pi r_b^2 l_b \, n_b \, \gamma_b m_e c^2$, where $r_b$ and $l_b$ are the beam radius and length respectively, while 
the maximal energy lost for target ionisation is $U_k=\pi r_b^2 \, \Delta z \, n_i \, U_k^\text{a}$, where $\Delta z$ is the target thickness and $U_k^\text{a}$ is the sum of the atomic ionization energies for the $k$ outermost electrons. For an isolated Strontium atom, we have\cite{NIST_ASD} $U_3^\text{Sr}\sim60~\text{eV},~U_9^\text{Sr}\sim658~\text{eV}$ corresponding approximately to $\xi_\text{low}$ and to $\xi^*$ respectively. 
Thus, the ratio between the maximal energy lost for target ionisation and the electron beam energy is: $U_k/U_e=\Delta z\,n_i\,U_k^a/(l_b\,n_b\gamma_bm_ec^2)$. 
We now consider different values for $n_b$ and $k$ while keeping constant the parameters $\Delta z=0.05$ cm, $l_b=10~\mu$m, $n_i^\text{Sr}=1.74\times10^{22}$ cm$^{-3}$ and $\gamma_b=4\times10^3$. 
For the electron beam densities relevant for this work $n_b\geq3\times10^{19}$, and we obtain: $U_3/U_e\lesssim10^{-3}$ and $U_9/U_e\lesssim10^{-2}$. Hence, in all relevant cases only a small fraction of the electron beam energy is lost to ionize the target. 
Here, we also point out that $Z_\text{eff}n_i>10^{23}$ cm$^{-3}$ and therefore the electron beam density ($\lesssim10^{21}$ cm$^{-3}$) accounts just for a very small correction to the target electron number density.  

We assign the values $n_e=Z_\text{eff}n_i-n_b$ to the free electron density and $\boldsymbol{v}_e=-\boldsymbol{v}_b\,n_b/n_e$ to their velocity in order to ensure the electric charge and the electric current density neutralisation at the beginning of each PIC simulation.  
The above initial condition is consistent with the EPOCH electromagnetic fields which are identically zero by default at $t_0=0$, and is justified by the fast generation of a return current within the charge relaxation time\cite{MillerChargedBeams,2001PhPl....8.1441F} $\tau_e\sim10^{-18}$ s. 
Also, we stress that at most only a small fraction of the beam energy is needed to generate the target electrons return current. In fact, the ratio between the target electron energy density $\Upsilon_e=n_em_ev_e^2/2$ and the electron beam energy density $\Upsilon_b=n_b\gamma_bm_ec^2$ is $\Upsilon=\Upsilon_e/\Upsilon_b=n_ev_e^2/(2n_b\gamma_bc^2)\approx n_b/(2n_e\gamma_b)$, where in the last approximated equality we used the relation $v_e=-v_bn_b/n_e\approx-cn_b/n_e$. 
Thus, by considering the numerical values corresponding to the setup described in this paragraph, we obtain $10^{-9}\lesssim\Upsilon\lesssim10^{-6}$.

\subsubsection*{Multiple scattering}

Multiple scattering\cite{Moliere:1948zz,PhysRev.89.1256} is the sequence of elastic Coulomb collisions a charged particle undergoes with the atoms of the target. 
The solution to the multiple scattering problem implemented in our code was obtained for the first time by Fermi and it was reported by Rossi and Greisen\cite{RevModPhys.13.240}. 
This simple model allows us to calculate the stochastic corrections to the trajectories of the electrons of the beam while they travel across the target. 
Such corrections are applied at each time step after the particle position and momentum are updated according to the particle equation of motion. 

The implementation in the code is as follows. Let us assume that the electron is located at the point $\boldsymbol{r}=(x,y,z)$ with momentum $\boldsymbol{p}=(p_{x},p_{y},p_{z})$. Since the electrons of the beam are ultra-relativistic and their velocity is almost collinear with the $z$-axis, the distance they travel after a time step $\Delta t$ is approximately $\Delta z\approx c\Delta t$, such that we can neglect the consequences of the multiple scattering on the electron longitudinal position $z$ and momentum $p_z$. 
Thus, the cumulative effect of the multiple scattering events which occurred along the path $\Delta z$ basically affects only the transverse electron displacements and the corresponding angular deviations\cite{Olive:2016xmw}:

\begin{equation}\label{deltaxy_thetaxy}
\Delta x=\frac{\Delta z\,\theta_0}{2}\left(\frac{g_1}{\sqrt{3}}+g_2\right)\;,\quad
\theta_x = g_2\,\theta_0\;,\quad
\Delta y=\frac{\Delta z\,\theta_0}{2}\left(\frac{g_3}{\sqrt{3}}+g_4\right)\;,\quad
\theta_y = g_4\,\theta_0\;,
\end{equation}
where $\theta_0=(13.6 \text{ MeV}/\beta c p)\sqrt{d\Delta z/X_0}$ is the root-mean-square of the Gaussian angular distribution, $p=|\boldsymbol{p}|$, $\beta=p/\sqrt{p^2+m_e^2c^2}$, and
$g_i$ with $i=1,\ldots,4$ are independent Gaussian random variables with mean zero and variance one. 
We stress that the necessary conditions for the applicability of the Fermi model, i.e. $\Delta x,\Delta y\ll\Delta z$ and $\theta_x,\theta_y\ll1$, are always fulfilled in our PIC simulations. 
Finally, the electron position and momentum are: $\boldsymbol{r}'=(x+\Delta x,y+\Delta y,z)$ and $\boldsymbol{p}'=R_y(\theta_x)R_x(\theta_y)\boldsymbol{p}$, where $R_x(\theta_y)$ ($R_y(\theta_x)$) corresponds to the rotation by an angle $\theta_y$ ($\theta_x$) around the $x$-axis ($y$-axis).

\subsubsection*{Bremsstrahlung}

An electron that is moving in a solid target generates bremsstrahlung radiation because it experiences the electronic and the nuclear electric fields.
The energy $\epsilon_\gamma$ of the photons generated via bremsstrahlung is obtained from the ultra-relativistic differential cross-section inclusive of Coulomb corrections in the complete screening limit \cite{RevModPhys.31.920}: 

\begin{equation}\label{dsigmabs_dk}
\frac{d\sigma_\text{bs}}{d\epsilon_\gamma}=\frac{4\alpha Z^2r_0^2}{\epsilon_\gamma}\left\{\left[1+\left(\frac{\epsilon_e-\epsilon_\gamma}{\epsilon_e}\right)^2-\frac{2}{3}\frac{\epsilon_e-\epsilon_\gamma}{\epsilon_e}\right]
c_0(Z)+\frac{1}{9}\frac{\epsilon_e-\epsilon_\gamma}{\epsilon_e}\right\}\;,
\end{equation}
where $\alpha=e^2/\hbar c \approx 1/137$ is fine structure constant, $r_0=e^2 / m_e c^2$ is the classical electron radius, $Z$ is the atomic number of the target, $\epsilon_e$ and $\epsilon_\gamma$ are the initial electron and emitted photon energies, respectively, $c_0=\ln\left(183/Z^{1/3}\right)-f_c(Z)$ and $f_c(Z)\approx0.925\,(\alpha Z)^2$. 

Due to the ultra-relativistic approximation, this formula is rigorously valid only when $\epsilon_\gamma$, $\epsilon_e$ and $\epsilon_e-\epsilon_\gamma$ are much larger than $m_e c^2$. 
However, by comparing the theoretical prediction given by equation (\ref{dsigmabs_dk}) for an ultra-relativistic electron with the corresponding tabulated reference values\cite{Seltzer1986345}, we verified that equation~(\ref{dsigmabs_dk}) holds with a few percent accuracy also for very low photon energies, where the bremsstrahlung emission probability is the highest.  
Furthermore, note that although equation~(\ref{dsigmabs_dk}) overestimates the emission probability near the region where $\epsilon_\gamma\sim \kappa_0=\epsilon_e-m_ec^2$, the probability of emission of such high-energy photons remains negligible for ultra-relativistic electrons. 

In order to compute the total probability for a bremsstrahlung event, we use the total bremsstrahlung cross section obtained by integrating equation~(\ref{dsigmabs_dk}):

\begin{equation}
\sigma_\text{bs}=\int_{\epsilon^*_\gamma}^{\kappa_0}d\epsilon_\gamma\;\frac{d\sigma_\text{bs}}{d\epsilon_\gamma}=4\alpha Z^2r_0^2\left\{\left(\frac{4\,c_0(Z)}{3}+\frac{1}{9}\right)\left[\ln\left(\frac{\kappa_0}{\epsilon^*_\gamma}\right)-\frac{\kappa_0-\epsilon^*_\gamma}{\epsilon_e}\right]+\frac{c_0(Z)}{2\epsilon_e^2}\left(\kappa_0^2-{\epsilon^*_\gamma}^2\right)\right\}\;,
\end{equation}
where the introduction of the lower integration limit $\epsilon^*_\gamma$ is motivated by the Landau-Pomeranchuk-Migdal (LPM) suppression\cite{LP1,LP2,1956PhRv..103.1811M}. 
Taking into account the electron beam energy and the target parameters adopted in this work, we set $\epsilon^*_\gamma=0.4\,m_ec^2\approx 200$~keV. 

\subsubsection*{Synchrotron radiation}

The energy $\epsilon_\gamma$ of the photon produced by an ultra-relativistic electron in the presence of external electromagnetic fields is derived from the spectral probability distribution in the magnetic bremsstrahlung limit\cite{Ritus1985,BaierBook}

\begin{equation}
dW(u)=\frac{\alpha m_e^2}{\sqrt{3}\hbar \epsilon_e}\frac{du}{(1+u)^3}
\left\{
\left[1+(1+u)^2\right]K_{2/3}\left(\frac{2u}{3\chi_e}\right)
-(1+u)\int_{2u/3\chi_e}^{\infty}dy\,K_{5/3}(y)
\right\}\;,
\end{equation}
where $u=\epsilon_\gamma/(\epsilon_e-\epsilon_\gamma)$, $\epsilon_e$ is the initial electron energy and

\begin{equation}\label{chi_def}
\chi_e=\frac{e\hbar \epsilon_e}{m_e^3c^5}\sqrt{\left(\boldsymbol{E}+\frac{\boldsymbol{v}\times\boldsymbol{B}}{c}\right)^2
-\left(\frac{\boldsymbol{v}\cdot\boldsymbol{E}}{c}\right)^2}\;,
\end{equation}
is the electron quantum parameter. 
Since $\chi_e\lesssim 2\times10^{-2}$ for the parameters adopted in this work, our results can also be understood and interpreted by employing the classical Larmor formula, which provides the electron instantaneous energy loss by emission of synchrotron radiation:

\begin{equation}\label{dE_dt}
\frac{dE}{dt}
=-\frac{2}{3}\frac{e^4}{m_e^4c^7}\epsilon_e^2
\left[\left(\boldsymbol{E}+\frac{\boldsymbol{v}\times\boldsymbol{B}}{c}\right)^2
-\left(\frac{\boldsymbol{v}\cdot\boldsymbol{E}}{c}\right)^2\right]
\approx-\frac{2}{3}\frac{e^4}{m_e^4c^7}\epsilon_e^2F_\perp^2
\;,
\end{equation}
where the transverse electromagnetic field is $F_\perp=\sqrt{(E_x-B_y)^2+(E_y+B_x)^2}$ and the ultra-relativistic approximation $\boldsymbol{v}\approx(0,0,c)$ for the electron velocity is employed in the last equality of equation~(\ref{dE_dt}). Moreover, the classical theory provides the following simple expression for the average photon energy in the case of synchrotron emission\cite{Jackson}:

\begin{equation}\label{ave_photon_ene}
\langle \epsilon_\gamma\rangle=
\frac{4}{5\sqrt{3}}\frac{e\hbar}{m_e^3c^5}\epsilon_e^2F_\perp\;.
\end{equation}

\subsection*{Diffusion}

Periodic boundary conditions are a satisfactory approximation as long as diffusion is negligible. 
While the longitudinal beam spreading can be safely neglected for ultra-relativistic beams, the lateral spreading becomes more important with increasing target thickness $z$. 
Thus, it is important to estimate the maximum target thickness $z_\text{max}$ that should be used in numerical computations performed with periodic boundary conditions, for a given beam radius. 
In view of the beam and the target parameters considered in our work, the transverse displacement due to the beam divergence is $\Delta r_{\text{div}}\approx8\times10^{-4}\,z$, while for multiple scattering\cite{Olive:2016xmw} we have $\Delta r_{\text{ms}}=4\times10^{-3}(d/X_0)^{1/2}z^{3/2}\approx1.9\times10^{-3}z^{3/2}$.
By setting $z_\text{max}=0.05$~cm we obtain $\Delta r_\text{diff}=\Delta r_{\text{div}}+\Delta r_{\text{ms}}\approx0.6$ $\mu$m that is adequate if the initial beam radius is larger than about $5$~$\mu$m.

\subsection*{Photon attenuation} 

Although high-energy photons are produced within a dense material in the presence of large electromagnetic fields, in this section we show that only a very small fraction of the generated photons can be absorbed (see Supplementary information for further details). 

First, the intensity $I(\epsilon_\gamma)$ after a monochromatic photon beam with energy $\epsilon_\gamma$ and initial intensity $I_0$ has travelled across a target with thickness $z$ and mass density $d$ is given by\cite{Olive:2016xmw} $I(\epsilon_\gamma)=I_0\exp\left[-zd\sigma_\text{tot}(\epsilon_\gamma)\right]$, where the total photon absorption cross section is $\sigma_\text{tot}=\sigma_\text{PE}+\sigma_\text{CS}+\sigma_\text{PP}$, where $\sigma_\text{PE}$, $\sigma_\text{CS}$ and $\sigma_\text{PP}$ are the cross sections for photoelectric effect (PE), Compton scattering (CS) and electron-positron pair production (PP), respectively. In particular, when $Z=38$ and $\epsilon_\gamma\geq \epsilon^*_\gamma=0.4~m_ec^2$, from tabulated cross sections we obtain\cite{XCOM} $\sigma_\text{tot}\lesssim0.2$~cm$^2$/g. As a consequence, by taking into account that the target mass density is $d_\text{Sr}\approx2.54$~g/cm$^3$ and the largest target thickness considered in this manuscript is $z_\text{max}=0.05$, it follows that $I/I_0\gtrsim0.97$ (see Supplementary information). 

Second, the probability for a photon with energy $\epsilon_\gamma$ to convert into an electron-positron pair in the presence of a transverse electromagnetic field with strength $F_\perp$, i.e. the Breit-Wheeler electron-positron pair production process, can be estimated from the value of the photon quantum parameter $\chi_\gamma= (\epsilon_\gamma/m_ec^2)(F_\perp/F_\text{cr})$, where $F_\text{cr}\approx4.4\times10^{13}$~esu is the QED critical field. The average number of electron-positron pairs generated over a distance $z$ can be estimated as: 
$N_{e^\pm}(\chi_\gamma,\epsilon_\gamma)=W_e(\chi_\gamma,\epsilon_\gamma)\,z/c$, where\cite{BaierBook} $W_e$ is the probability of electron-positron pair creation per unit time. Since $\chi_\gamma\lesssim2\times10^{-2}$ for the parameters adopted in this work, $W_e$ increases monotonically with increasing photon energy when $\epsilon_\gamma\leq\epsilon^\text{max}_\gamma$, where $\epsilon^\text{max}_\gamma=\gamma_bm_ec^2+\Delta E_b=2.5~\text{GeV}$, and $\Delta E_b\sim0.5~\text{GeV}$ accounts for the electron beam energy spread. As a consequence, considering that $N_\pm\lesssim10^{-8}$ for $\epsilon^\text{max}_\gamma$ and $z_\text{max}=0.05$~cm, the Breit-Wheeler process is negligible for all photon energies considered here (see Supplementary information).

\section*{Data availability}

The data that support the plots and findings of this paper are available from the corresponding author upon reasonable request.

\bibliography{references.bib}{}

\section*{Acknowledgements}

The authors would like to thank Dr.~Naveen~Kumar for valuable discussions and suggestions.

\section*{Author contributions}

A.B. initially conceived the research project with input from M.T., configured and carried out the simulations, generated the figures and wrote the bulk of the manuscript. M.T. implemented the routines accounting for multiple scattering, bremsstrahlung and synchrotron radiation into the particle-in-cell code EPOCH. A.B. and M.T. discussed the physics, and analysed and interpreted the results of the simulations. C.H.K. supervised the project. All authors contributed to the preparation of the manuscript.

\section*{Competing financial interests}

The authors declare no competing financial interests.

\end{document}